\documentclass[journal]{IEEEtran}
\usepackage{amsmath,amsfonts}
\usepackage{algorithmic}
\usepackage{array}
\usepackage{stfloats}
\usepackage{url}
\usepackage{graphicx}
\usepackage{multirow}
\usepackage{float}
\usepackage{cite}
\usepackage[table,xcdraw]{xcolor}
\usepackage{lettrine}
\usepackage{bm}
\usepackage{booktabs}
\usepackage{subfigure}

\begin{document}

\title{Deep Learning for Spectrum Prediction in Cognitive Radio Networks: State-of-the-Art, New Opportunities, and Challenges}

\author{Guangliang~Pan,~\IEEEmembership{Member,~IEEE,}
	David~K. Y.~Yau,~\IEEEmembership{Senior~Member,~IEEE,}\\
	Bo~Zhou,~\IEEEmembership{Member,~IEEE,}
	and Qihui~Wu,~\IEEEmembership{Fellow,~IEEE}		
\thanks{This work was supported in part by the National Natural Science Foundation of China under Grants 62231015 and 62201255. (\textit{Corresponding authors: David~K. Y.~Yau, Qihui Wu})}
\thanks{G. Pan, B. Zhou, and Q. Wu are with the College of Electronic and Information Engineering, Nanjing University of Aeronautics and Astronautics, Nanjing, 211106, China (e-mail: \{glpan2020, b.zhou, wuqihui\}@nuaa.edu.cn).}
\thanks{D. Yau is with the Pillar of Information Systems Technology and Design, Singapore University of Technology and Design, Singapore 487372 (e-mail: david\_yau@sutd.edu.sg).}
}

\maketitle

\begin{abstract}
Spectrum prediction is considered to be a promising technology that enhances spectrum efficiency by assisting dynamic spectrum access (DSA) in cognitive radio networks (CRN). Nonetheless, the highly nonlinear nature of spectrum data across time, frequency, and space domains, coupled with the intricate spectrum usage patterns, poses challenges for accurate spectrum prediction. Deep learning (DL), recognized for its capacity to extract nonlinear features, has been applied to solve these challenges. This paper first shows the advantages of applying DL by comparing with traditional prediction methods. Then, the current state-of-the-art DL-based spectrum prediction techniques  are reviewed and summarized in terms of intra-band and cross-band prediction. Notably, this paper uses a real-world spectrum dataset to prove the advancements of DL-based methods. Then, this paper proposes a novel intra-band spatiotemporal spectrum prediction framework named ViTransLSTM. This framework integrates visual self-attention and long short-term memory to capture both local and global long-term spatiotemporal dependencies of spectrum usage patterns. Similarly, the effectiveness of the proposed framework is validated on the aforementioned real-world dataset. Finally, the paper presents new related challenges and potential opportunities for future research.  
\end{abstract}


\section{Introduction}\label{sec1}
\lettrine[lines=2]{T}{he} wide application of 5G mobile communication and Internet of Things (IoT) network technology has made the number of wireless devices grow by hundreds of millions every year, which puts a huge load on the already scarce radio spectrum resources. The International Telecommunication Union (ITU) allocates spectrum resources to these wireless devices in a static manner, restricting their communication to licensed frequency bands. Consequently, licensed bands experiencing an inundation of wireless devices face congestion issues and degradation in the quality of service \cite{10595399}. In contrast, other licensed bands, such as those designated for broadcasting TVs or analogue cellular telephony, possess abundant available spectrum that is regrettably underutilized.

To address these problems, cognitive radio network (CRN)-based dynamic spectrum access (DSA) is considered an effective solution. It enables wireless devices, referred to as secondary users (SUs), to opportunistically access the unused licensed spectrum as long as harmful interference to primary users (PUs) is limited. This access method can achieve better spectrum efficiency and higher system capacity, thereby alleviating the shortage of spectrum resources. The key step in DSA is to accurately obtain the spectrum state by spectrum sensing in cognitive radio (CR). Spectrum sensing techniques in various frequency bands include energy detection, matched filtering ($<$ 1 GHz), feature detection (sub-6 GHz), compressed sensing, and machine learning (millimeter wave and terahertz) techniques. However, practical hardware constraints limit spectrum sensing, including factors such as limited time delay, energy constraints, and sensing scopes. Fortunately, spectrum prediction can obtain unknown spectrum occupancy information in advance by capturing potential correlation patterns in the measured spectrum data to help CUs quickly access idle bands \cite{ding2017spectrum}. Herein, spectrum prediction focuses on providing future spectrum state, while spectrum sensing emphasizes obtaining the current state. Therefore, spectrum prediction has emerged as a research hotspot. 

However, spectrum prediction has some tricky challenges. This paper divides the existing work into intra-band prediction and cross-band prediction. In intra-band prediction, spectrum measurement is highly nonlinear due to diverse wireless devices, varied spectrum services, and influences from both the external environment and internal device interference. This nonlinearity spans time, frequency, and space, making learning multidimensional nonlinear features challenging. Cross-band prediction is the prediction of the future spectrum of the target band by using a few spectrum measurements in the target band (typically caused by spectrum security, device deployment, and hardware damage, etc) and abundant spectrum measurements in relevant bands. Effectively leveraging the substantial data from relevant bands and the limited data from the target band to achieve cross-band prediction poses a formidable challenge.

Recently, inspired by the stunning breakthroughs that deep learning (DL) has achieved in computer vision and natural language processing, DL has also been harnessed in spectrum prediction \cite{9625078, 8998400, 10064355}. DL can directly extract nonlinear usage patterns from spectrum measurements and integrate various types of networks such as convolutional neural networks (CNN) and recurrent neural networks (RNN) to capture multidimensional correlations (see \cite{10064355} for specific correlations). In cross-band prediction, DL enhances the performance of target band prediction by increasing the number of samples in the target band using generative adversarial networks (GAN) and transferring useful knowledge from related bands. 

Based on these motivations, this paper provides a thorough overview of DL-based spectrum prediction in CRN. Specifically, we first demonstrate the advantages of applying DL by conducting a comparative analysis with traditional spectrum prediction methods. Furthermore, we review existing works and summarize them into two routes: intra-band prediction  \cite{9625078, 8998400, 10064355, 10039050, 9296309, 9664805} and cross-band prediction  \cite{10122825, 9339826, 9020298, 9657212}, with detailed information available in Table \ref{tab2}. Secondly, to enhance spatiotemporal spectrum prediction performance, we propose a novel spatiotemporal spectrum prediction framework named ViTransLSTM by combining visual self-attention and long short-term memory (LSTM). This framework introduces a shifted-window-based visual self-attention mechanism into the LSTM gate structure and memory cells, enabling it to capture both local and global spatiotemporal dependencies by collaboratively accumulating memory with each gate structure. We validate the effectiveness of the proposed framework on a real-world spectrum dataset. Finally, we provide future challenges related to spectrum prediction in real-world wireless systems and provide DL-based research directions for these challenges. The basic concepts and technical terms mentioned in this paper are summarized in Table \ref{tab1}.
\begin{figure}
	\centerline{\includegraphics[width=88mm,height=60mm]{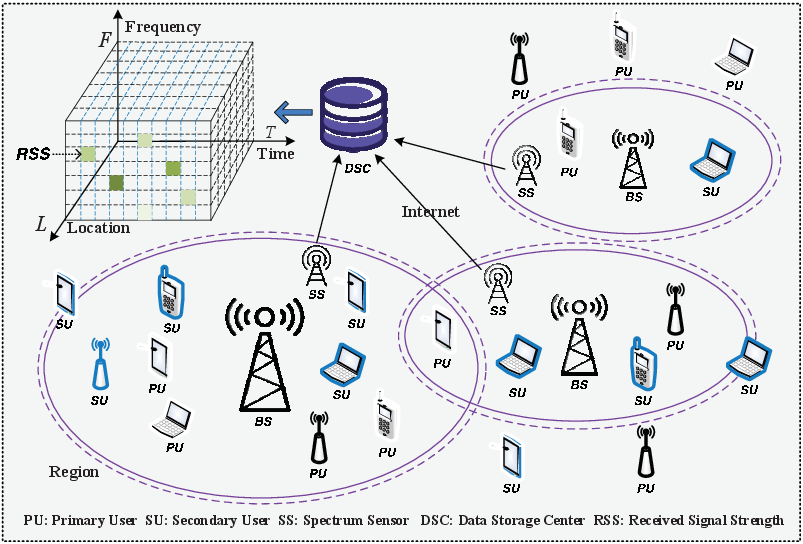}}
	\caption{System model.}
	\label{Fig:framework}
\end{figure}

\section{Why DL for Spectrum Prediction?}\label{sec2}
As shown in Fig. \ref{Fig:framework}, we consider a CRN consisting of several PUs and SUs. The spectrum sensors (SSs), which are sparsely distributed in the region of interest (RoI), transmit the received signal strength (RSS) over the Internet to the data storage center (DSC). Referencing the upper left corner of Fig. \ref{Fig:framework}, spectrum measurement can be divided into time, frequency, and space domains. DL enables spectrum prediction by learning the underlying temporal/spectral/spatial correlations in historical spectrum measurement. Below, we give the reasons for using DL by comparing it with traditional methods.

\textit{AR Model-Based Methods.} Influenced by user behavior and radio equipment activity, latent time patterns exist in the time dimension of spectrum, encompassing periodicity and trends. To explore these patterns, AR and moving average (MA) models use parameter estimation to analyze the influence of past spectrum series on the current moment, aiming to better capture temporal patterns for predicting future spectrum states \cite{ding2017spectrum}. The ARMA model, which combines AR and MA, is then used to further enhance analytical capabilities. Subsequently, a more advanced AR integrated MA (ARIMA) model is proposed to eliminate trends or seasonal effects by introducing differential operations \cite{ding2017spectrum}. However, these models assume that the relationship between the past and future values of the spectrum series is linear. In real-world scenarios, many spectrum series exhibit nonlinear behavior, and these models may struggle to capture such patterns. In contrast, DL does not require any prior assumptions and can automatically extract complex nonlinear patterns, thereby achieving accurate spectrum prediction.

\textit{Traditional Machine Learning (ML)-Based Methods.} Traditional ML methods, including support vector machine (SVM), hidden Markov model (HMM), and Bayesian inference have been utilized for spectrum prediction \cite{ding2017spectrum}. Authors of \cite{ding2017spectrum} introduce that SVM predicts spectrum mobility based on design features, HMM predicts spectrum occupancy with state transition matrices, and Bayesian inference estimates channel quality using a posterior distribution. However, SVM usually requires manual selection and crafting of features. HMM comes with limitations in terms of context modeling, relying solely on previous observations and struggling to capture long-term dependencies. Bayesian methods face constraints due to their reliance on specific probabilistic distributions, limited access to prior knowledge, and sensitivity to the selection of prior distributions. Moreover, these models are only used to capture features in the temporal dimension of the spectrum, ignoring the frequency and spatial dimensions. In contrast, DL takes an end-to-end approach, eliminating the need for manual feature engineering. DL-models are highly flexible and can adapt to a wide range of tasks without substantial modifications. Moreover, DL-models can capture long-term spectrum patterns to support accurate prediction.

\section{Learning Nonlinearities for Intra-Band Spectrum Prediction}\label{sec3}
\subsection{Time-Frequency Spectrum Prediction}
Various frequency bands are allocated for diverse spectrum services, encompassing applications such as broadcasting TV, GSM900 (both uplink and downlink), ISM, GSM1800 (both uplink and downlink), and similar services. The interplay of factors such as the number of spectrum devices, user mobility, and patterns of usage imparts unique temporal and spectral correlations to each spectrum service. The intricacies of this multifaceted time-frequency correlation are notably nonlinear, presenting a challenge for conventional methodologies. To overcome this challenge, the application of DL-models becomes pertinent. These models prove effective in capturing the nuanced time-frequency correlations by directly assimilating spectrum correlations from measurements across distinct frequency bands, subsequently encapsulating this information in a parameterized format within the neural network architecture.

\begin{table*}
	\centering
	\renewcommand\arraystretch{1.7}		
	\caption{Introduction to Basic Concepts and Technical Terminology.}\label{tab1}
	\begin{tabular}{>{\columncolor[HTML]{DAE8FC}}m{2.8cm}
			>{\columncolor[HTML]{ECF4FF}}m{12cm}}
		\hline  
		\textbf{Concepts and Terms} & \textbf{Introduction}  \\
		\hline 
		CRN  &  An intelligent wireless communication system that dynamically adapts its operations to the surrounding environment by sensing and utilizing available spectrum resources efficiently. \\ 
		Intra-band Prediction  &  It involves inferring the future spectrum evolution of a given band based on its past spectrum measurements over a certain period. \\
		Cross-band Prediction & It involves inferring the future spectrum evolution of a related but different band using the past spectrum measurements of both the given band and the related band.  \\
		Deep Learning &  A subset of machine learning that utilizes neural networks with multiple layers to automatically extract and learn complex patterns and representations from large amounts of data. \\
		Temporal patterns & The potential trends and cycles in spectrum data over time (such as proximity and diurnal cycles). \\
		Tensor Completion  &  It estimates the missing or unobserved entries in a partially observed tensor (a multi-dimensional array). \\
		Deep TL  &  It combines DL with TL techniques to leverage pre-trained neural networks for efficiently adapting knowledge from one domain to improve performance in a different but related domain. \\
		Multimodal Learning & An approach that integrates and processes information from multiple data modalities, such as text, images, and audio, to achieve a more comprehensive understanding and enhanced predictive performance. \\
		Knowledge Drift & The phenomenon where the relevance or applicability of knowledge transferred from a source domain diminishes or becomes inconsistent when applied to a dynamically changing target domain.  \\
		\hline
	\end{tabular}		
\end{table*}

As an attempt, LSTM was employed to learn the temporal correlations of multiple spectrum channels. Gao et al. \cite{9625078} utilized LSTM and attention mechanisms to design a sequence-to-sequence model for achieving multi-channel, multi-step spectrum prediction. Supervised learning is adopted for the model optimization. Once the model is adequately trained, supervised learning transitions to the prediction phase. In this stage, only a specific length of historical spectrum data is input into the prediction model to predict the future spectrum state without any prior information. To further improve predictive performance, Yu et al. \cite{8998400} combined CNN and gated recurrent unit (GRU) (a variant of the LSTM) networks to design a model called DCG for spectrum availability prediction. Recognizing the local correlations among different channels within the same time slot and regional correlations across multiple time slots, Yu et al. employed one-dimensional convolution to delve into the occupancy patterns of local channels and two-dimensional convolution to explore the occupancy patterns of regional channels. Subsequently, GRU was leveraged to encapsulate both short-term and long-term temporal dependencies within the data processed by the dual CNN.

Given the extensive range of frequency bands utilized by wireless devices, the amalgamation of vast spectrum data inherently results in high-dimensional data. Simultaneously, there is a heightened demand for the model's proficiency in extracting features. To tackle this challenge, Pan et al. \cite{10064355} initially employed stacked autoencoders (SAE) to diminish the dimensionality of the high-dimensional spectrum data while automatically extracting features without disrupting its internal temporal and frequency relationships. The extracted features were then fed into a CNN and bidirectional LSTM (Bi-LSTM) fusion network to grasp the frequency and time dependencies. However, due to the constraints of the Bi-LSTM gating structure, it demonstrated excellent short-term predictive performance but relatively weaker long-term predictive performance. To surmount this limitation, Pan et al. \cite{10039050}, drawing inspiration from the Transformer architecture, devised a long-term spectrum prediction method named Autoformer-CSA. Autoformer-CSA employs a self-attention mechanism, integrating series spatial and channel attention modules with autocorrelation mechanisms. This enables the model to allocate varying degrees of attention to different positions in the spectrum sequence, effectively capturing the long-term trends and seasonal characteristics of spectrum data.

\begin{figure}
	\centerline{\includegraphics[width=72mm,height=62mm]{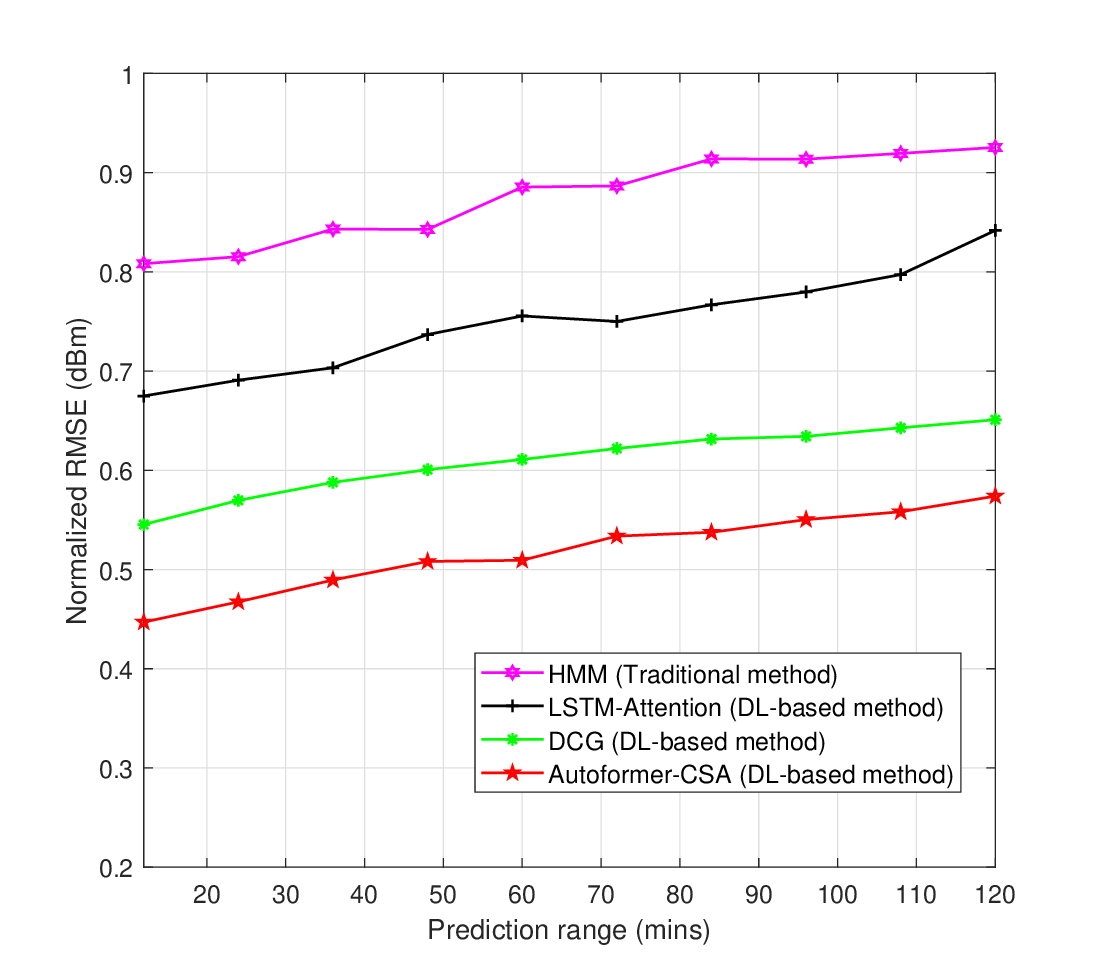}}
	\caption{RMSE comparison of various spectrum prediction schemes. All schemes are executed on a PC featuring an Intel(R) Xeon(R) E5-2698 v4 CPU @ 2.20GHz, NVIDIA Tesla V100 GPU 32GB graphics card, and 256GB RAM, utilizing PyTorch 2.1.0 with the Python programming language. The DL-based schemes use mean squared error (MSE) as the loss function, with 20 training epochs, a batch size of 32, and an early stopping patience of 6.}
	\label{Fig2}
\end{figure}

Fig. \ref{Fig2} presents a comparison of the root mean square error (RMSE) performance between the DL-based approach and the traditional HMM prediction method across various prediction ranges. The dataset utilized for these comparative experiments is derived from real-world spectrum data collected by sensors deployed in the city center of Madrid, Spain (obtained via the Electrosense open API, https://electrosense.org/). The scanned frequency range spans 600-640 MHz with a 2 MHz sampling interval. Data collection took place from June 1, 2021, to June 8, 2021, with a sampling interval of 1 mins. The division of the training set, validation set, and test set followed a 5:1:1 ratio. Fig. \ref{Fig2} shows that the RMSE errors of the DL-based prediction methods are significantly lower than those of the traditional HMM-based prediction method. For instance, at a prediction range of 96 mins, LSTM with attention, DCG, and Autoformer-CSA exhibit 14.66\%, 30.58\%, and 39.77\% higher RMSE performance, respectively, compared to HMM.
\begin{table*}
	\centering
	\renewcommand\arraystretch{1.6}		
	\caption{DL for radio spectrum prediction: state-of-the-art.}\label{tab2}
	\begin{tabular}{>{\columncolor[HTML]{DAE8FC}}m{2.5cm}
			>{\columncolor[HTML]{ECF4FF}}m{0.5cm}
			>{\columncolor[HTML]{DAE8FC}}m{3cm}
			>{\columncolor[HTML]{ECF4FF}}m{2cm}
			>{\columncolor[HTML]{DAE8FC}}m{7cm}}
		\hline
		\multicolumn{5}{c}{\cellcolor[HTML]{EFEFEF} \textbf{Learning Nonlinearities for Intra-Band Spectrum Prediction}}\\
		\hline  
		\textbf{Research routes} & \textbf{Refs.} & \textbf{Learning methods} & \textbf{DL models} & \textbf{Key features} \\
		\hline  
		Time-frequency spectrum prediction  & \cite{9625078}  & Supervised learning & LSTM with Attention & Learn multi-channel temporal correlations \\
		& \cite{8998400}  & Supervised learning & CNN-GRU & Learn temporal-spectral correlations \\
		& \cite{10064355} & Unsupervised learning and supervised learning & SAE-Bi-LSTM & Dimensionality reduction, automatic feature extraction, and temporal-spectral correlation learning \\
		& \cite{10039050} & Supervised learning & Transformer  & Learn the long-term dependence of spectrum series \\
		Time-frequency-space spectrum prediction & \cite{9296309} & Supervised learning & PredRNN  & Learn temporal-spectral-spatial correlations  \\
		& \cite{9664805} & Supervised learning & CNN-ResNet  & Use the ResNet to reduce gradient disappearance and learn spatio-temporal correlations \\
		\hline
		\multicolumn{5}{c}{\cellcolor[HTML]{EFEFEF} \textbf{Learning Enhancement for Cross-Band Spectrum Prediction}}\\
		\hline
		\textbf{Research routes} & \textbf{Refs.} & \textbf{Learning methods} & \textbf{DL models} & \textbf{Key features} \\
		\hline
		DGAN for cross-band spectrum prediction & \cite{10122825} & Unsupervised learning and supervised learning & GAN & Improve cross-band prediction performance with data enhancement \\
		& \cite{9339826} & Unsupervised learning and transfer learning & DCGAN & Improve cross-band prediction performance with data enhancement and fine-tuning of model parameters \\
		DTL for cross-band spectrum prediction & \cite{9020298} & Transfer learning & DTL  & Use naive parameter transfer to improve the performance of target band prediction  \\
		& \cite{9657212} & Transfer learning & DTL   & Use weighted parameter transfer to improve the performance of target band prediction \\
		\hline
	\end{tabular}		
\end{table*}

\subsection{Time-Frequency-Space Spectrum Prediction}
Unlike traditional time-frequency prediction, DL-based spatiotemporal spectrum prediction faces two key challenges: reconstructing spectrum map data for large RoIs and capturing heterogeneous time-frequency-space correlations.

To overcome these challenges, Li et al. \cite{9296309} initially delved into historical data collected from a network of sparsely distributed spectrum sensors. They employed an inverse-distance spatial interpolation method to extrapolate the spectrum map across the entire RoI. They utilized predictive recurrent neural network (PredRNN) to discern the spatial-temporal correlations embedded in the spectrum map. Employing three identical PredRNN components, they modeled the temporal proximity, daily cycle, and weekly trend of the spectrum image. Ultimately, Li et al. used parameter matrices to integrate the features extracted from these three components. This refined approach to feature extraction has proven effective in significantly enhancing the model's predictive performance.

According to the path loss model in \cite{9664805}, it's clear that wireless signals follow exponential decay with distance. As the average received power at nearby locations tends to be similar, there's a higher spatial correlation. In simpler terms, only sensors close to the unsensed area can provide signal power information that has a certain correlation with the data in the unsensed area. Therefore, Ren et al. \cite{9664805} went a step further and employed a neighboring spatial interpolation method to estimate the spectrum map for the entire RoI. This method is compared with a tensor completion, and the proposed approach achieves the minimum error rate at a low sparsity level. Note that in addition to the above methods there are Kriging, kernel-based, and improved tensor completion methods. Ren et al. proposed a spatiotemporal spectrum prediction method combining CNN and ResNet to capture spectrum matrix spatiotemporal features, using skip connections to improve performance and prevent gradient vanishing.

\section{Learning Enhancement for Cross-Band Spectrum Prediction}\label{sec4}
In real-world spectrum services, certain frequency bands, like those used for cellular mobile communication, can gather a substantial amount of spectrum data. Conversely, other bands, such as military bands, may have only limited available spectrum data. For frequency bands with abundant labeled data, we can directly employ DL-models for training and prediction. However, in scenarios where data is scarce, training DL-models becomes challenging, resulting in less-than-optimal predictive performance. To tackle this challenge, there are currently two methods: one involves increasing the amount of trainable data for the target band to support model training, while the other involves transferring knowledge from related bands with ample available data to the target prediction network. Below, we detail how DL drives both methods.
\subsection{DGAN for Cross-Band Spectrum Prediction}
A deep GAN (DGAN) consists of a generator and a discriminator. The GAN's aim is to train the generator network to produce realistic data, while the discriminator network distinguishes between generated and real data. These two networks engage in mutual adversarial training, propelling the learning and improvement of the model. Consequently, GAN can be employed to boost the quantity of trainable data for the target band.

Peng et al. \cite{10122825} introduced a spectrum data conversion GAN to generate realistic data for the target band. Initially, they used Fréchet inception distance (FID) to measure differences between the target band and other bands, pinpointing the source band with the least difference from the target predictive band. Peng et al. then crafted a generator using a blend of CNN and LSTM to transform data from the source band to the target predictive band. Subsequently, the prediction model utilized both the transformed data and the original data from the target band for training and prediction.

It's essential to note that this method doesn't directly generate the target prediction data but leverages highly similar data from other bands. This is because traditional GANs have specific requirements for the quantity of target samples, and insufficient target samples can lead to GAN instability. Additionally, this approach reduces the dependency of the GAN on existing data from the target band. Lin et al. \cite{9339826} similarly utilized FID to identify a source band with high similarity to the target band. They then devised a deep convolutional GAN (DCGAN) for pre-training using the source band. Subsequently, transfer learning was employed to fine-tune the pre-trained model onto the target band, generating data with minimal differences. Finally, the generated data for the target band, along with the original data, was used for training and prediction in a residual prediction model.

\subsection{DTL for Cross-Band Spectrum Prediction}
Transfer learning (TL) aims to distill knowledge from one or multiple source tasks and apply that knowledge to related but different target tasks. Building upon this TL concept, deep transfer learning (DTL) has emerged as a secondary approach to tackle this challenge. DTL efficiently achieves cross-band spectrum prediction by transferring pre-trained models from the source band to the target task.

Lin et al. \cite{9020298} implemented cross-band prediction through a naive TL approach by transferring a prediction model comprised of LSTM units. To ensure positive TL, Lin et al. analyzed the similarity between source and target bands and used dynamic time warping to measure the similarity of each frequency point between source and target bands. Subsequently, Lin et al. employed a transfer component analysis method, utilizing maximum mean discrepancy as the distance measurement, based on the marginal distribution of features to obtain features with strong transferability. 

For an additional performance enhancement, Li et al. proposed a transfer time-frequency fusion attention network (T-TF$^2$AN) to achieve cross-band spectrum prediction. The primary challenge in applying TL for TF$^2$AN stems from the gaps between source and target bands, resulting from variations in spatial, temporal, or spectral domains. Li et al. addressed these challenges through weighted TL, adaptively discovering and transferring shared knowledge while mitigating the negative impact of specific domain patterns from the source spectrum. The core components of the designed weighted TL are the shared pattern learner and the loss with adaptive weights. The shared pattern learner assigns higher weights to source spectrum data similar to the target spectrum, incorporating more shared patterns. Adaptive weights are determined by appropriately weighting the loss from the source domain, diminishing the risk of negative transfer and effectively enhancing the model's performance on the target domain dataset. It's noteworthy that the T-TF$^2$AN model not only considers cross-band prediction but also takes into account spatial and temporal transferability.
\begin{figure*}
	\centerline{\includegraphics[width=178mm,height=65mm]{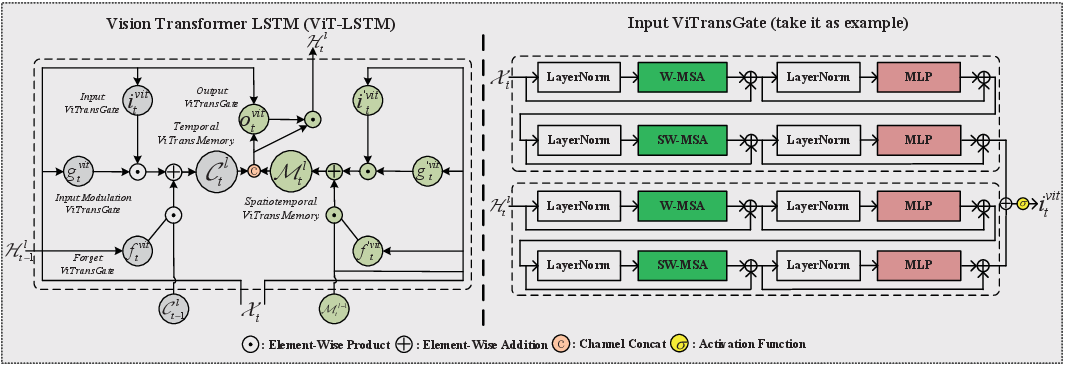}}
	\caption{The overall structure of the ViT-LSTM.}
	\label{Fig5}
\end{figure*}

\section{A ViTransLSTM Spatiotemporal Spectrum Prediction Framework}
\subsection{Preblem Description}
Assuming that there are $S$ sparsely distributed spectrum sensors in a RoI divided into $M (\text{rows}) \times N (\text{columns})$ grids to measure the radio signal power in the area. Considering the hardware cost, it is not feasible to deploy sensors at every location. Therefore, we adopt the commonly used inverse-distance spatial interpolation method to fill in the missing signal power at unmeasured locations. These measurements at any given time step $t$ can be represented as a tensor $\mathcal{X}_{t} \in \mathbb{R}^{C \times M \times N}$, where the measurements are mapped to three channels ($C=3$) using a common Jet colormap. The spatiotemporal spectrum prediction problem involves predicting a spectrum sequence of the most probable length-$K$ time steps in the future based on a spectrum sequence of historical length-$J$ time steps, including measurements at the current time step, represented as
\begin{equation}\label{Eq_1}
\begin{aligned}
\hat{\mathcal{X}}_{t+1}&, \cdots, \hat{\mathcal{X}}_{t+K} = \\
	&\mathop {\text{arg max}} \limits_{\mathcal{X}_{t+1}, \cdots, \mathcal{X}_{t+K}} p(\mathcal{X}_{t+1}, \cdots, \mathcal{X}_{t+K} \vert \mathcal{X}_{t-J+1}, \cdots, \mathcal{X}_{t}).
\end{aligned}
\end{equation}
\subsection{ViTransLSTM Framework}
In spatiotemporal spectrum prediction, most works adopt convolutional structures to capture spatial correlations, such as PredRNN used in \cite{9296309}. However, due to the limitations of convolutional kernel size, convolutional structures may struggle to capture global information. The fixed parameter-sharing mechanism in convolution operations makes it challenging for the model to handle spatiotemporal correlations at different locations. Compared to convolutional structures, vision Transformers (ViTs) excel at capturing spatial correlations due to their self-attention-based global learning patterns. Inspired by this, we integrate ViT and PredRNN to propose an extended spatiotemporal spectrum prediction framework called ViTransLSTM. The innovation of this framework lies in replacing the core component ST-LSTM in PredRNN with our designed ViT-LSTM component. The overall structure of the ViT-LSTM component is shown on the left of Fig. \ref{Fig5}. 

As a variant of ST-LSTM, spatiotemporal ViT-LSTM consists of two sets of gate structures: a standard temporal ViTransMemory and a spatiotemporal ViTransMemory. In the standard temporal ViTransMemory, all the inputs $\mathcal{X}_{t-J+1}$, $\cdots$, $\mathcal{X}_{t}$, hidden states $\mathcal{H}_{t-J+1}$, $\cdots$, $\mathcal{H}_{t}$, cell outputs $\mathcal{C}_{t-J+1}$, $\cdots$, $\mathcal{C}_{t}$, and gates, i.e., input ViTransGate $i^{vit}_t$, input-modulation ViTransGate $g^{vit}_t$, forget ViTransGate $f^{vit}_t$ are 3D tensors in $\mathbb{R}^{C \times M \times N}$. In the spatiotemporal ViTransMemory, all the inputs $\mathcal{X}_{t-J+1}$, $\cdots$, $\mathcal{X}_{t}$, hidden states $\mathcal{M}^{l-1}_{t-J+1}$, $\cdots$, $\mathcal{M}^{l-1}_{t}$, cell outputs $\mathcal{M}^{l}_{t-J+1}$, $\cdots$, $\mathcal{M}^{l}_{t}$, and gates ${i^{'}}^{vit}_t $, ${g^{'}}^{vit}_t$, ${f^{'}}^{vit}_t$, $o^{vit}_t$ are also 3D tensors in $\mathbb{R}^{C \times M \times N}$. The equations of ViT-LSTM in the $l$-th layer are shown as follows:  
\begin{equation}\label{vit}
	\begin{aligned}
		&g^{vit}_t =\text{tanh}(\text{vit}(\mathcal{X}_t)+\text{vit}(\mathcal{H}_{t-1}^{l})+b_g) \\
		&i^{vit}_t =\sigma(\text{vit}(\mathcal{X}_t)+\text{vit}(\mathcal{H}_{t-1}^{l})+b_i) \\
		&f^{vit}_t =\sigma(\text{vit}(\mathcal{X}_t)+\text{vit}(\mathcal{H}_{t-1}^{l})+b_f) \\
		&\mathcal{C}_t^l =f^{vit}_t \odot  \mathcal{C}_{t-1}^l + i^{vit}_t \odot g^{vit}_t\\
		&{g^{'}}^{vit}_t =\text{tanh}(\text{vit}(\mathcal{X}_t) + \text{vit}(\mathcal{M}_{t}^{l-1}) + b^{'}_g) \\
		&{i^{'}}^{vit}_t =\sigma(\text{vit}(\mathcal{X}_t) + \text{vit}(\mathcal{M}_{t}^{l-1}) + b^{'}_i) \\
		&{f^{'}}^{vit}_t =\sigma(\text{vit}(\mathcal{X}_t) + \text{vit}(\mathcal{M}_{t}^{l-1}) + b^{'}_f) \\
		&\mathcal{M}_t^l ={f^{'}}^{vit}_t \odot  \mathcal{M}_{t}^{l-1} + {i^{'}}^{vit}_t \odot {g^{'}}^{vit}_t\\
		&o^{vit}_t =\sigma(\text{vit}(\mathcal{X}_t) + \text{vit}(\mathcal{H}_{t-1}^{l}) + \text{vit}(\mathcal{C}_{t}^{l}) + \text{vit}(\mathcal{M}_{t}^{l}) + b_o) \\
		&\mathcal{H}_t^l = o^{vit}_t \odot \text{tanh}(\text{linear}([\mathcal{C}_t^l, \mathcal{M}_t^l])),
	\end{aligned}
\end{equation}
where $\sigma$ is the sigmoid activation function, $\ast$ is the convolution operator, $\odot$ is the Hadamard product, $[\cdot, \cdot]$ is the operation of concatenating two tensors, and $\text{vit}(\cdot)$ is two consecutive Swin Transformer \cite{liu2021swin} blocks (see Fig. \ref{Fig5}, right), which includes a standard multi-head self-attention module with non-overlapping window (W-MSA), a MSA module with non-overlapping shifted window (SW-MSA), and a feed-forward network (FFN, is a 2-layer multilayer perceptron (MLP), with Gaussian error linear unit (GELU) non-linearity in between) following each MSA module. Layer normalization (LN) is applied before each MSA module and FFN, and a residual connection is applied after each module.

As shown in (\ref{vit}), unlike ST-LSTM, ViT-LSTM introduces two new designs: \textit{ViTransGate} and \textit{ViTransMemory}. Specifically, 1) \textit{ViTransGate:} This design replaces the convolutional networks in all gate structures of the original ST-LSTM with ViT (i.e., Swin Transformer). Taking the input ViTransGate $i^{vit}_t$ (see Fig. \ref{Fig5}, right) as an example, it is obtained by feeding $\mathcal{X}_t$ and $\mathcal{H}_{t-1}^{l}$ into the $\text{vit}(\cdot)$ module, followed by the activation function. 2) \textit{ViTransMemory:} From (\ref{vit}), this design feeds the tamporal memory unit $\mathcal{C}_t^l$ and the spatiotemporal memory unit $\mathcal{M}_t^l$ separately into the $\text{vit}(\cdot)$.

Compared to the original ST-LSTM, the advantages of ViT-LSTM include: 1)  The ViTransGates in ViT-LSTM capture local and global spectral pattern features of the spectrum map by computing self-attention within the standard window and the shifted window, respectively. In contrast, the original ST-LSTM can only capture local features through convolution operations, which may lead to the loss of critical spatial spectrum pattern features. 2) The ViTransMemory also employs the self-attention mechanism to integrate spectral information from different positions, capturing more diverse feature representations to enhance memory. Additionally, it works in conjunction with LSTM to store long-range spatiotemporal spectrum dependencies. In contrast, convolution operations may exhibit biases toward local patterns, potentially leading to the loss of specific memory features.

\subsection{Experimental Results}
To validate the effectiveness of the proposed ViTransLSTM, we use spectrum data collected by four sensors via the Electrosense open API (https://electrosense.org/) as the dataset. This dataset is reconstructed into $64 \times 64$ spectrum maps based on inverse distance spatial interpolation of the data from these four sensors (sensor ID: $\text{test}\_\text{yago}\_3$, $\text{test}\_\text{yago}$, $\text{rack}\_3$, and $\text{rack}\_2$). The data collection period spans from Jun. 1 to Jun. 3, 2021, with a frequency band of 610 MHz and a sampling interval of 1 mins. The dataset (a total of 3200 samples) is divided into training, validation, and test sets in a ratio of 4:1:1. In the ViTransLSTM, the window size of the $\text{vit}(\cdot)$ network is $4\times4$, the patch size is $4\times4$, and the number of heads is 8. We select recently proposed spectrum prediction models, including ConvLSTM, PredRNN \cite{9296309}, and the upgraded version PredRNN V2 \cite{9749915}, as baseline comparisons. We run all the experiments on a PC with 3.30 GHz Intel Core i9-10940X CPU, NVIDIA GTX 3090Ti graphic, and 64 GB RAM using the Pytorch 1.8.0. All models are trained using the Adam optimizer with a starting learning rate of 0.001. The training process is stopped after 1000 iterations. Unless otherwise specified, batch size is 4.

\begin{figure*}
	\centering
	\subfigure[Average MSE with predicting 10 frames]{
		\label{Fig:MSE}\includegraphics[width=74mm,height=66mm]{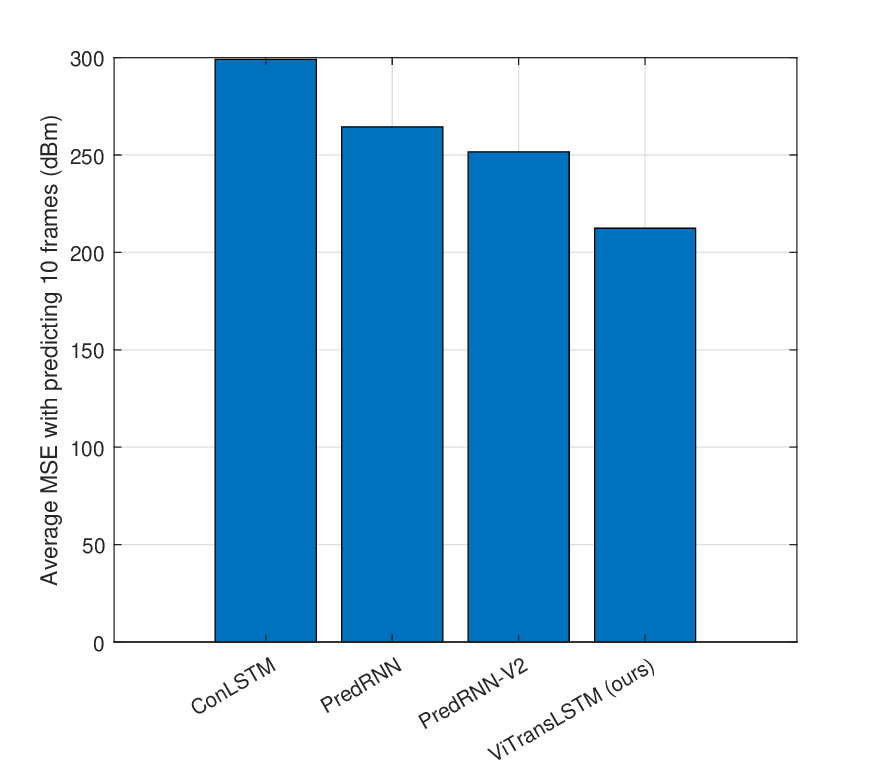}}
	\subfigure[Average PSNR with predicting 10 frames]{
		\label{Fig:PSNR}\includegraphics[width=74mm,height=66mm]{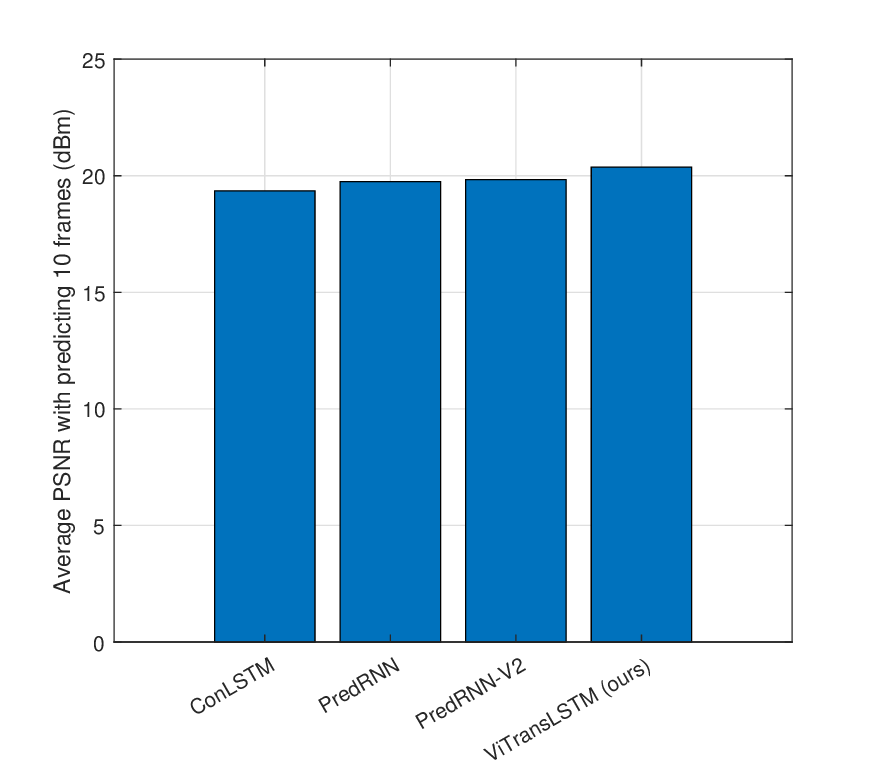}}
	\caption{Comparison results of the average MSE and PSNR between ViTransLSTM and baselines under the prediction of 10 frames.}
	\label{msepsnr}	
\end{figure*}

Under the setup of input-10-predict-10, Fig. \ref{msepsnr} presents the comparison results of the proposed ViTransLSTM and all baselines in terms of average MSE and PSNR. As shown in Fig. \ref{msepsnr}, the proposed ViTransLSTM outperforms all baselines across all evaluation metrics. For example, in Fig. \ref{Fig:MSE}, the average MSE of ViTransLSTM is 28.99\%, 19.70\%, and 15.58\% lower than that of ConvLSTM, PredRNN, and PredRNN V2, respectively. Similarly, in Fig. \ref{Fig:PSNR}, the average PSNR of ViTransLSTM is 5.29\%, 3.13\%, and 2.72\% higher than that of ConvLSTM, PredRNN, and PredRNN V2, respectively. These results demonstrate that the proposed ViTransLSTM, which integrates visual self-attention with LSTM, can effectively capture deeper spatiotemporal correlation features of spectrum usage patterns, thereby achieving more accurate predictions.

\section{Research Challenges and Opportunities}
\begin{figure*}
	\centerline{\includegraphics[width=164mm,height=94mm]{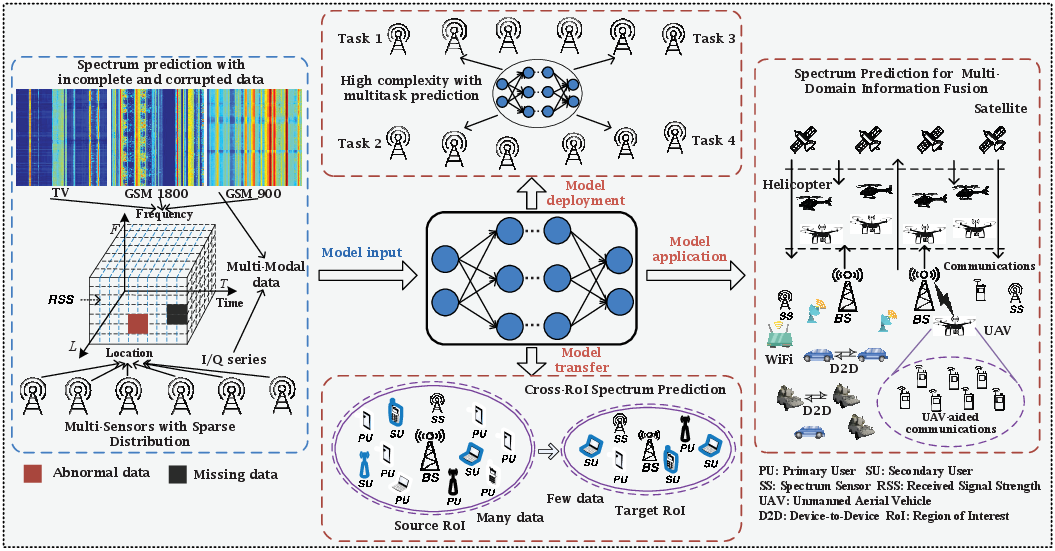}}
	\caption{Summary of research challenges for DL-based spectrum prediction.}
	\label{Fig4}
\end{figure*}

Although DL has achieved beneficial performance gains in spectrum prediction, there are still numerous open challenges for further study, as summarized in Fig. \ref{Fig4} at a glance.

\subsection{Spectrum Prediction with Incomplete and Corrupted Data}
Owing to interference from spectrum measuring equipment, signal propagation environments, and other variables, spectrum measurements may contain missing and abnormal data. This compromise in data quality amplifies the complexity of modeling. In an initial exploration, tensor completion is used to address this challenge. However, tensor completion algorithms struggle to estimate nonlinear missing data accurately. Fortunately, the diffusion model uses a stepwise de-noising approach to grasp the distribution of nonlinear spectrum data, resulting in the generation of high-quality spectrum data that is robust to deletions and anomalies.

\subsection{Multi-Modal Integration Assisted Spectrum Prediction}
Despite achieving impressive accuracy, existing DL-based spectrum prediction methods rely only on uni-modal data, highlighting the need for a robust DL-based multi-modal fusion framework to integrate features from each modality (such as spectrum occupancy series and spectrogram). To the best of our knowledge, deep multi-modal learning has yet to be applied in the field of spectrum prediction, despite its successful implementation in other domains such as weather and traffic predictions. However, a significant challenge involves the necessity to assign relative weights to different modalities. A viable approach could use a multi-objective optimization algorithm to balance the optimal weight of each modal by treating each weight as an optimization objective.

\subsection{High Complexity with Multi-Tasking Spectrum Prediction}
Multi-tasking spectrum prediction is often considered to meet the customization needs of massive spectrum users. However, when various DL-based spectrum prediction models are deployed in multiple spectrum prediction tasks, the intrinsic complexity of the models often necessitates substantial computational resources for training. To address this issue, it is recommended to customize lightweight DL models for spectrum prediction to achieve a flexible balance between performance and complexity through pruning and knowledge distillation techniques. Further, the lightweight model uses distributed learning \cite{9562559} for each task node to improve computing efficiency.

\subsection{Cross-RoI Spatiotemporal Spectrum Prediction}
Unlike cross-band spectrum prediction, cross-region spatiotemporal spectrum prediction extends into the time-frequency-space dimension. TL addresses this challenge, with cross-region TL going beyond time-frequency knowledge to include spatial information. This introduces complexity, leading to a challenge known as knowledge drift. As more knowledge transfers, the problem of knowledge drift becomes more severe. Enhancing transfer learning efficiency and spectrum prediction accuracy in the target region is a significant challenge. To address this, importance weight TL reweights source domain knowledge to improve model performance. Furthermore, cross-region cross-band spectrum prediction based on DTL presents an even more formidable problem. The relevance of the target domain to the source domain is lower compared to the previous challenge, intensifying the knowledge drift problem in transfer learning. Refinement TL emerges as a solution, involving the classification of transfer knowledge to improve transfer efficiency.

\subsection{Spectrum Prediction for Multi-Domain Information Fusion}
The electromagnetic spectrum is evolving, integrating space, air, and ground communication modes like vehicle-to-vehicle (V2V), unmanned aerial vehicle (UAV)-assisted, air-to-ground, and satellite communication. These modes create complex spectrum usage patterns. Joint spectrum prediction for space, air, and ground is challenging due to multi-domain information fusion, addressed by spectrum information semanticization. Additionally, traditional-DL struggles in dynamic wireless systems, requiring frequent retraining for changing data distributions, consuming time and memory. Deep reinforcement learning offers a solution by training intelligent agents to perform tasks and derive optimal strategies from experience. In highly dynamic wireless systems, where the environment constantly changes, the agent receives feedback through a reward mechanism and makes rational decisions.

\section{Conclusion}\label{sec5}
We have commenced by elaborating on the motivation of applying DL in spectrum prediction. Firstly, DL is eminently suitable for extracting the complex nonlinear features encountered. Secondly, the adaptive fitting capability of DL enables parameter knowledge fine-tuning required by the predicted band change. Based on these motivations, DL is proposed for extracting both the intrinsic nonlinear time-frequency- and time-frequency-space-domain features, thereby inspiring short/long-term spectrum predictions. Further, DL was employed for data enhancement of target prediction bands and knowledge transfer of source bands to achieve cross-band spectrum prediction. Then, a framework combining visual self-attention and LSTM, named ViTransLSTM, is proposed to achieve high-precision spatiotemporal spectrum prediction. We have validated the effectiveness of the proposed framework on a real-world spectrum dataset. Finally, we have provided the future associated challenges and potential opportunities of applying DL to spectrum prediction. 

\bibliographystyle{IEEEtran}
\bibliography{myref}

\begin{thebibliography}{10}
\providecommand{\url}[1]{#1}
\csname url@samestyle\endcsname
\providecommand{\newblock}{\relax}
\providecommand{\bibinfo}[2]{#2}
\providecommand{\BIBentrySTDinterwordspacing}{\spaceskip=0pt\relax}
\providecommand{\BIBentryALTinterwordstretchfactor}{4}
\providecommand{\BIBentryALTinterwordspacing}{\spaceskip=\fontdimen2\font plus
\BIBentryALTinterwordstretchfactor\fontdimen3\font minus
  \fontdimen4\font\relax}
\providecommand{\BIBforeignlanguage}[2]{{%
\expandafter\ifx\csname l@#1\endcsname\relax
\typeout{** WARNING: IEEEtran.bst: No hyphenation pattern has been}%
\typeout{** loaded for the language `#1'. Using the pattern for}%
\typeout{** the default language instead.}%
\else
\language=\csname l@#1\endcsname
\fi
#2}}
\providecommand{\BIBdecl}{\relax}
\BIBdecl

\bibitem{10595399}
H.~Wang, Q.~Wu, and W.~Chen, ``Movable antenna enabled interference network:
  Joint antenna position and beamforming design,'' \emph{IEEE Wireless Commun.
  Lett.}, vol.~13, no.~9, pp. 2517--2521, 2024.

\bibitem{ding2017spectrum}
G.~Ding, Y.~Jiao, J.~Wang, Y.~Zou, Q.~Wu, Y.-D. Yao, and L.~Hanzo, ``Spectrum
  inference in cognitive radio networks: Algorithms and applications,''
  \emph{IEEE Commun. Surv. Tutor.}, vol.~20, no.~1, pp. 150--182, 2017.

\bibitem{9625078}
Y.~Gao, C.~Zhao, and N.~Fu, ``Joint multi-channel multi-step spectrum
  prediction algorithm,'' in \emph{Proc. IEEE Veh. Technol. Conf.}, 2021, pp.
  1--5.

\bibitem{8998400}
L.~Yu, Y.~Guo, Q.~Wang, C.~Luo, M.~Li, W.~Liao, and P.~Li, ``Spectrum
  availability prediction for cognitive radio communications: A {DCG}
  approach,'' \emph{IEEE Trans. Cogn. Commun. Netw.}, vol.~6, no.~2, pp.
  476--485, 2020.

\bibitem{10064355}
G.~Pan, Q.~Wu, G.~Ding, W.~Wang, J.~Li, F.~Xu, and B.~Zhou, ``Deep stacked
  autoencoder based long-term spectrum prediction using real-world data,''
  \emph{IEEE Trans. Cogn. Commun. Netw.}, vol.~9, no.~3, pp. 534--548, 2023.

\bibitem{10039050}
G.~Pan, Q.~Wu, G.~Ding, W.~Wang, J.~Li, and B.~Zhou, ``An autoformer-{CSA}
  approach for long-term spectrum prediction,'' \emph{IEEE Wireless Commun.
  Lett.}, vol.~12, no.~10, pp. 1647--1651, 2023.

\bibitem{9296309}
X.~Li, Z.~Liu, G.~Chen, Y.~Xu, and T.~Song, ``Deep learning for spectrum
  prediction from spatial–temporal–spectral data,'' \emph{IEEE Commun.
  Lett.}, vol.~25, no.~4, pp. 1216--1220, 2021.

\bibitem{9664805}
X.~Ren, H.~Mosavat-Jahromi, L.~Cai, and D.~Kidston, ``Spatio-temporal spectrum
  load prediction using convolutional neural network and {R}es{N}et,''
  \emph{IEEE Trans. Cogn. Commun. Netw.}, vol.~8, no.~2, pp. 502--513, 2022.

\bibitem{10122825}
C.~Peng, R.~Zhu, M.~Zhang, and L.~Wang, ``Cross-band spectrum prediction
  algorithm based on data conversion using generative adversarial networks,''
  \emph{China Commun.}, vol.~20, no.~10, pp. 136--152, 2023.

\bibitem{9339826}
F.~Lin, J.~Chen, G.~Ding, Y.~Jiao, J.~Sun, and H.~Wang, ``Spectrum prediction
  based on {GAN} and deep transfer learning: A cross-band data augmentation
  framework,'' \emph{China Commun.}, vol.~18, no.~1, pp. 18--32, 2021.

\bibitem{9020298}
F.~Lin, J.~Chen, J.~Sun, G.~Ding, and L.~Yu, ``Cross-band spectrum prediction
  based on deep transfer learning,'' \emph{China Commun.}, vol.~17, no.~2, pp.
  66--80, 2020.

\bibitem{9657212}
K.~Li, C.~Li, J.~Chen, Q.~Zhang, Z.~Liu, and S.~He, ``Boost spectrum prediction
  with temporal-frequency fusion network via transfer learning,'' \emph{IEEE
  Trans. Mobile Comput.}, vol.~22, no.~6, pp. 3209--3223, 2023.

\bibitem{liu2021swin}
Z.~Liu, Y.~Lin, Y.~Cao, H.~Hu, Y.~Wei, Z.~Zhang, S.~Lin, and B.~Guo, ``Swin
  transformer: Hierarchical vision transformer using shifted windows,'' in
  \emph{Proc. IEEE Int. Conf. Comput. Vis. (ICCV)}, 2021, pp. 10\,012--10\,022.

\bibitem{9749915}
Y.~Wang, H.~Wu, J.~Zhang, Z.~Gao, J.~Wang, P.~S. Yu, and M.~Long, ``Predrnn: A
  recurrent neural network for spatiotemporal predictive learning,'' \emph{IEEE
  Trans. Pattern Anal. Mach. Intell.}, vol.~45, no.~2, pp. 2208--2225, 2023.

\bibitem{9562559}
M.~Chen, D.~Gündüz, K.~Huang, W.~Saad, M.~Bennis, A.~V. Feljan, and H.~V.
  Poor, ``Distributed learning in wireless networks: Recent progress and future
  challenges,'' \emph{IEEE J. Sel. Areas Commun.}, vol.~39, no.~12, pp.
  3579--3605, 2021.

\end{thebibliography}

\end{document}